"\pdfoutput=1"
\documentclass[twocolumn,prb,showpacs]{revtex4}
\usepackage{graphicx}
\usepackage{dcolumn}
\usepackage{bm}
\makeatletter

\newcommand{\Rmnum}[1]{\expandafter\@slowromancap\romannumeral #1@}
\makeatother

\begin{document}

\title{Direct experimental evidence of multiferroicity in nanocrystalline Zener Polaron ordered manganites}

\author{Vinay Kumar Shukla}
\email{vkshukla@iitk.ac.in}

\author{Soumik Mukhopadhyay}
\email{soumikm@iitk.ac.in}
\affiliation{Department of Physics, Indian Institute of Technology, Kanpur 208 016, India}
\author{Kalipada Das}
\author{A. Sarma}
\author{I. Das}
\affiliation{Saha Institute of Nuclear Physics, Kolkata 700 064, India}


\begin{abstract}
We discuss the particle size driven tunability of the coexistence of ferromagnetism and ferroelectricity in Pr$_{0.67}$Ca$_{0.33}$MnO$_{3}$ (PCMO) with the help of x-ray diffraction (XRD), magnetization, impedance spectroscopy and remanent polarization measurements. The remanent polarization measurements using `Positive Up Negative Down' (PUND) method clearly prove the existence of ferroelectricity in PCMO with phase separation between Zener Polaron (ZP) ordered \textit{P2$_{1}$nm} and disordered \textit{Pbnm} structures. We also find that the ferroelectric response is enhanced in nanocrystalline samples so long as ZP ordering is not destroyed while the long range antiferromagnetic ordering at low temperature in bulk system is replaced by ferromagnetic correlations in nanoparticles.  The conclusion, that by reducing the crystallite size, it might be possible to make ferromagnetism and ferroelectricity coexist near room temperature, should be generally applicable to all ZP ordered manganites.
[Published: Phys. Rev. B 90, 245126 (2014)]
\end{abstract}


\maketitle

\section{Introduction}

The ongoing debate on the dielectric properties of charge-ordered manganites is centered not only about the origin of spontaneous electric polarization but to experimentally establish the existence of the proposed ferroelectric ordering. Several theoretical models have been put forward: first, the static long-range ordering
of bond-centered (BC) `Zener polarons' (ZP)~\cite{Aladine} whereby one $e_{g}$ electron is shared by two equivalent Mn$^{3.5+}$ ions and the accompanying dimerisation results in the non-centrosymmetric \textit{P2$_1$nm} space group leading to spontaneous axial electric polarization~\cite{Wu, Jooss}; second, the coexistence of classic checkerboard type `site-centered' (SC) charge ordering and ZP/BC ordering could lead to the breaking of inversion symmetry causing axial as well as non-axial polarization~\cite{Khomskii, Efremov, Brink}; third, finite dislocation in a spin density wave, commensurate with respect to the lattice, can give rise to ferroelectric instability without breaking the inversion symmetry in the magnetic structure~\cite{Betouras}; fourth, strong electron-electron interactions combined with the Jahn-Teller (JT) lattice distortions can cause a canting instability of its antiferromagnetic (AFM) ground state and the resulting non-collinear magnetic ordering induces ferroelectricity of purely electronic origin~\cite{Giovannetti}.

Colizzi \textit{et al.}~\cite{Colizzi} has shown using advanced first principle calculations that Pr$_{0.5}$Ca$_{0.5}$MnO$_{3}$ consists of an electrically and magnetically polarized ground state having non axial polarization, similar in line with Khomskii's picture of co-existing SC and BC phases. The calculated non zero components of Berry phase electric polarization have both ionic and electronic origin. Yamauchi \textit{et al.}~\cite{Yamauchi} by using DFT simulations calculated two non zero components of total polarization, one of which is due to coexistence of BC-SC mechanisms and other is due to Heisenberg exchange interaction between parallel/antiparallel spin sites thus reaffirming the previous work of Colizzi \textit{et al.}~\cite{Colizzi}. Recently a simple physically intuitive picture on the origin of ZP ordering was presented~\cite{Barone} in which oxygen buckling caused by the tilting of the MnO$_{6}$ octahedra drives the system toward a ZP instability and hence spontaneous electric polarization. The ZP dimerised phase depends crucially on the
competition between double exchange, electron correlation, and the electron phonon coupling. The
more distorted the system is, the more effective the Coulomb and JT interactions are in stabilizing the ZP phase. Despite initial controversies~\cite{Grenier, Goff} the presence of ZP ordering in Pr$_{1-x}$Ca$_{x}$MnO$_{3}$ is now conclusively established~\cite{Aladine, Wu, Jooss}.

From the experimental point of view it has been difficult to establish directly the existence of ferroelectricity in charge ordered manganites due to the finite conductivity and the associated problem of leakage. Lopes \textit{et al.}~\cite{Lopes} studied the electric field gradient (EFG) across phase diagram of Pr$_{1-x}$Ca$_{x}$MnO$_{3}$  and observed a new phase transition in temperature dependence of EFG which was interpreted as evidence of electric polarization due to charge ordering. Jooss \textit{et al.} studied electrically polarized domains with electron microscopy in the Lorenz mode, the origin of which was attributed to ZP ordering~\cite{Jooss}. The domains were stable at room temperature where no magnetic ordering was observed. There have also been a few experimental studies attributing the dielectric anomaly near the charge ordering temperature to ferroelectric ordering or electronic phase separation~\cite{Mercone, Freitas}. In this article we not only attempt at exploring the interconnection between structural, magnetic and dielectric properties of Pr$_{0.67}$Ca$_{0.33}$MnO$_{3}$ (PCMO) but also the first direct measurement of remanent electric polarization of such systems both in bulk and nanocrystalline form.

During the last decade or so several experimental studies regarding the influence of particle size reduction on charge ordering and associated structural/electronic phase coexistence have been published~\cite{Das, Raju, Fang, Zhang, Chai, Lu, Samantaray, Wang, Sarkar, Giri, Tonogai, Jirak, Hill, Ping, Giri1, Biswas1, Biswas2, SM1, SM2, Kalipada1, Kalipada2}. And there is a general consensus now that with particle size reduction long range charge ordering is suppressed. Although a holistic understanding of the phenomenon is still lacking, the suppression of charge ordering in nanocrystalline manganites is generally attributed to a number of factors: 1) enhanced surface pressure on crystal structure leading to reduced orthorhombic distortion and unit cell volume and hence increased e$_{g}$ electron band width. This results in weakening of the charge-ordering and its eventual collapse 2) spin-canting in the anti-ferromagnetic (AFM) structure due to exchange biasing with surface spin configuration leading to weakening of charge order, etc. Interestingly, there are a few studies which claim that even though long range charge ordering is destroyed with particle size reduction, short range correlation persists~\cite{Zhou, Zhou1}.

All these plausible scenarios which lead to suppression of long range charge ordering can actually be bracketed under two major categories related to the general consequences of particle size reduction in any crystal: first, the surface and interface effects due to enhanced surface to volume ratio and second, the changes in lattice symmetry. Incidentally, both can have profound influence on the ferroelectric response as well. One of the primary objectives of this study is to investigate the influence of particle size reduction on the ferroelectric response in `charge-ordered' manganites since this opens up an opportunity to directly establish any correlation that may exist between long range charge ordering and ferroelectricity.

\section{Experiments}
Nanocrystalline Pr$_{0.67}$Ca$_{0.33}$MnO$_{3}$ samples were prepared by standard sol-gel method with Pr$_{6}$O$_{11}$, CaCO$_{3}$, and MnO$_{2}$ as starting compounds having purity of 99.99\% and subsequent heating at different temperatures (starting from 700$^\circ$ to 1000$^\circ$). The bulk polycrystalline sample was prepared by subjecting a single phase nanocrystalline sample to heat treatment at $1400^\circ$C for 36 hours. X-ray diffraction $\theta-2\theta$ scans at room temperature suggest that all the samples are of single phase. Using Scherrer's formula over the XRD line width broadening, the average crystal size for the samples was estimated.

In addition to the bulk sample, we discuss the results for 3 representative nanocrystalline samples of average crystallite size ${\sim}$ 32 nm, 55 nm and 80 nm, respectively. For convenience the bulk sample henceforth will be referred to as PCMO1, while the nanocrystalline samples as PCMO2 (average crystallite size 80 nm) and PCMO3 (average crystallite size 55 nm) and PCMO4 (average cystallite size 32 nm), respectively. The particle size distribution, crystallinity and chemical composition were studied using Transmission Electron Microscopy (TEM), selected area electron diffraction (SAED) and Energy Dispersive X-ray Spectroscopy (EDX), respectively. The oxygen stoichiometry was ascertained using Thermo-gravimetric analysis (TGA). The macroscopic magnetic measurement such as temperature and magnetic field dependence of magnetization was done using a Quantum Design vibrating sample magnetometer. The powder x-ray diffraction (XRD) measurements over the temperature range 10-300 K were carried out using Rigaku-TTRAX-III diffractometer having 18 kW rotating anode X-ray source. The temperature variation as well as the frequency dependence (over a broad range from 10Hz to 100MHz) of the real and imaginary part of dielectric constant were investigated with the help of a Novocontrol Alpha-A impedance analyzer. We complemented the results from impedance spectroscopy with effective dielectric constant and remanent polarization measurement using the `Positive-Up-Negative-Down' (PUND) method~\cite{Naganuma} with the help of a Radiant Premier II ferroelectric tester. The short range magnetic correlation was studied using Electron Spin Resonance (ESR) spectroscopy in a Bruker EMX spectrometer at the X band (9.46 GHz) in the temperature range 110-300 K. For impedance spectroscopy and for remanent polarization measurement we have used silver paste as electrodes on both sides of the sample in the form of circular disc of diameter 6 mm and thickness 1 mm.

\section{Results and discussion}

\subsection{Structural and magnetic characterization}

\begin{figure}
\includegraphics[width=8.5cm]{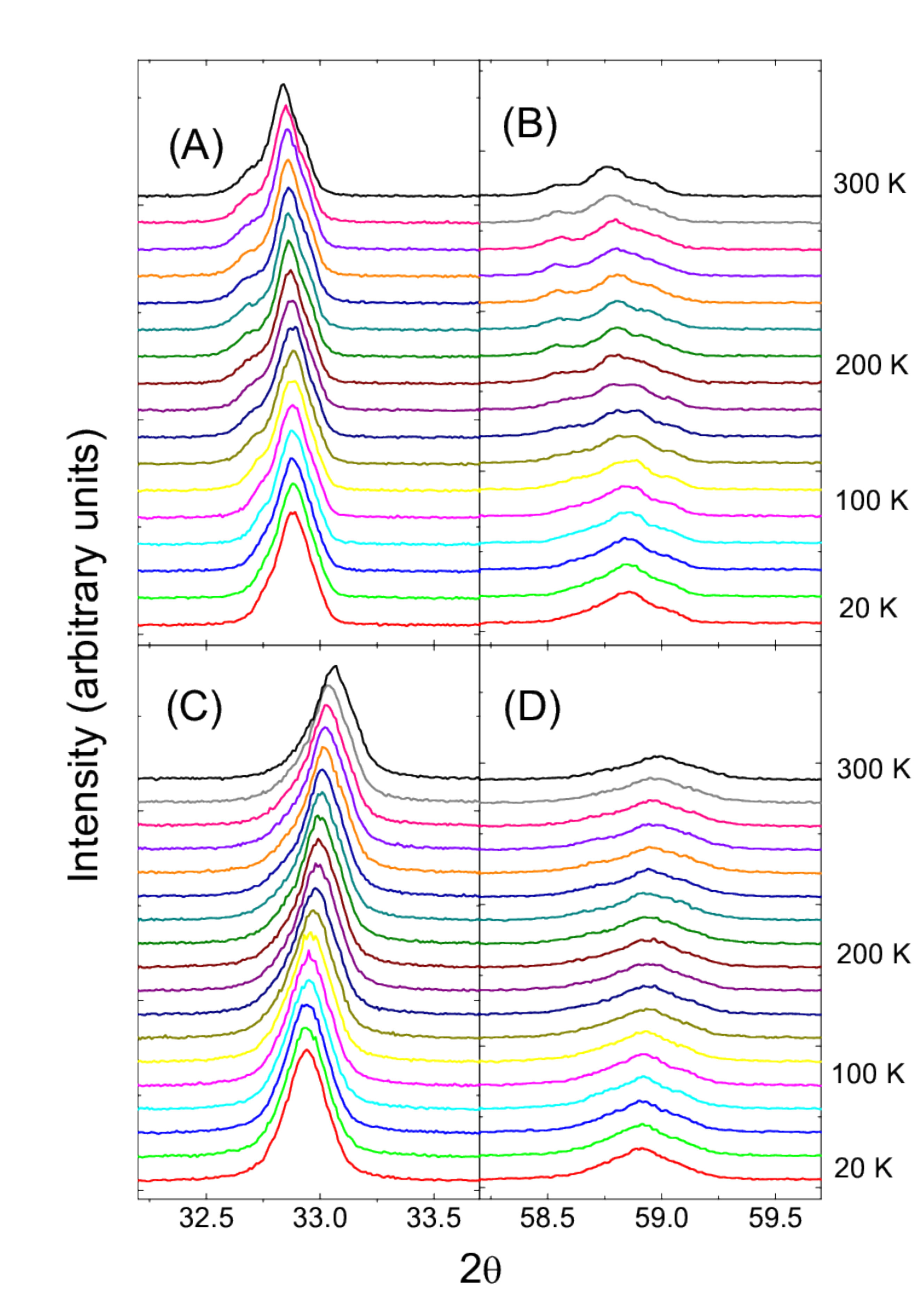}\\
\caption{Comparison of XRD $\theta-2\theta$ scan of selected Bragg peaks at different temperatures from 20K to 300K for (A, B) bulk PCMO1 and (C,D) nanocrystalline PCMO3 of average crystallite size $55$ nm. For PCMO1, the appearance of additional Bragg peaks with increasing temperature indicating structural phase coexistence is clearly discernible.}\label{fig:XRD}
\end{figure}

\begin{figure}
\includegraphics[width=8.5cm]{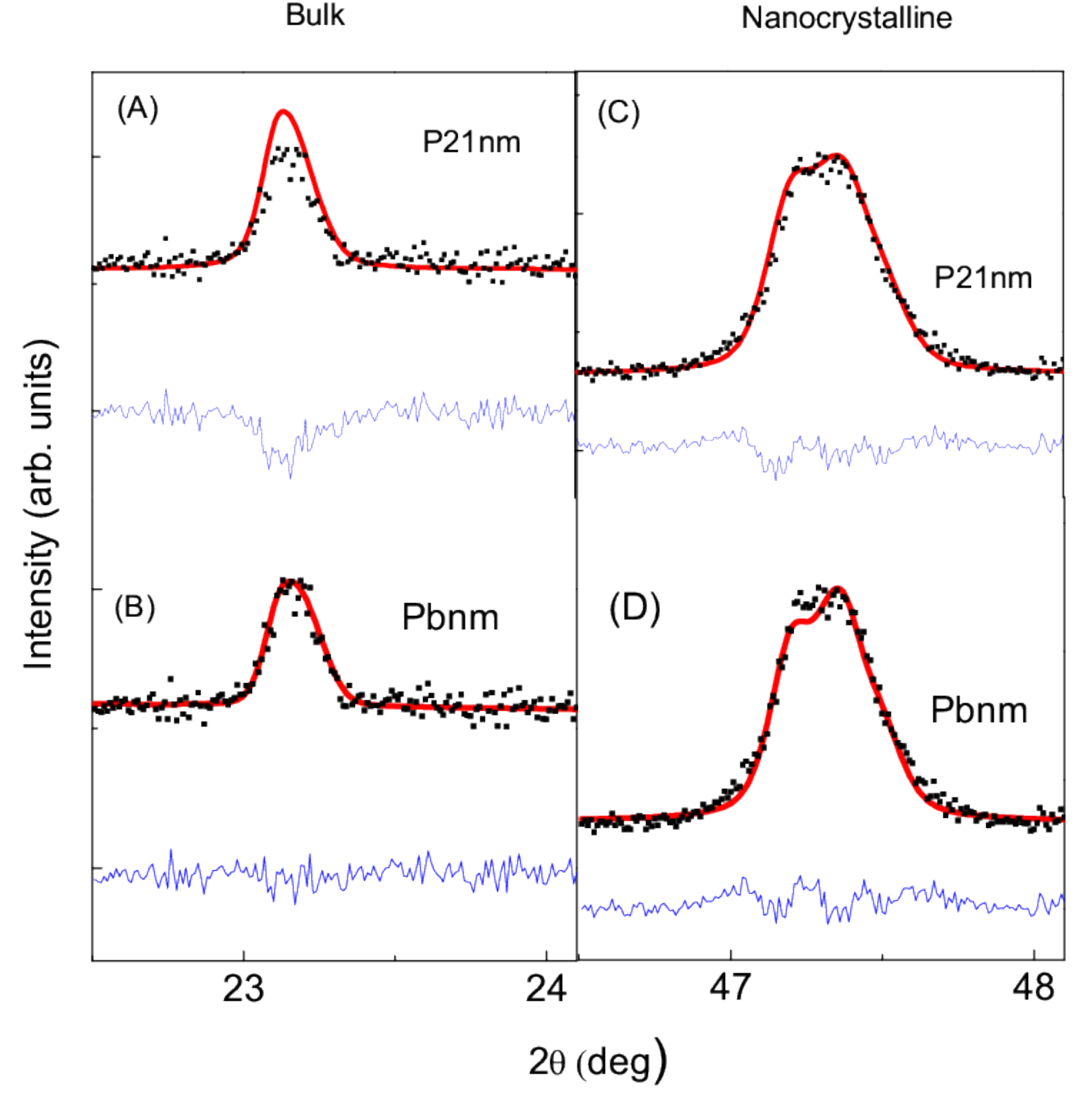}\\
\caption{Rietveld refinement analysis for bulk PCMO1 and nanocrystalline PCMO3 sample at 300K: for bulk, XRD data are fitted assuming (A)\textit{P2$_1$nm} space group and (B)\textit{Pbnm} space group and for PCMO3, we assume (C)\textit{P2$_1$nm} space group and (D)\textit{Pbnm} space group. It turns out that for the Bulk sample, \textit{Pbnm} is the most suitable space group at room temperature. For PCMO3, it is difficult to distinguish between different space group symmetries.}\label{fig:spacegroup}
\end{figure}

The determination of space group symmetry using conventional macroscopic powder XRD $\theta-2\theta$ scan (with the help of Rietveld refinement analysis using full prof suite) is a little tricky because of the existence of twin domains in PCMO. There are a few reports which even question the existence of ZP ordered state in PCMO based on Resonant XRD analysis~\cite{Grenier, Goff}. Besides, there are several space groups such as \textit{Pbnm, P11m, P2$_1$nm, P112$_1$/m, P112$_1$/b, P2$_1$nb, P11b}, etc. which are very close to each other in terms of symmetry properties and it is hard to distinguish as to which space group fits the given experimental data unambiguously the best, with the effect of surface disorder and line width broadening in nanoparticles making matter even more complicated. Nonetheless the temperature dependence of the powder diffraction scan shows development of additional Bragg peaks with increasing temperature in case of bulk sample signalling a change in space group symmetry while no such phenomenon is observed in case of nanoparticles (Fig~\ref{fig:XRD}). We find that room temperature symmetry of bulk sample is centrosymmetric \textit{Pbnm} (Fig.~\ref{fig:spacegroup})in agreement with previous reports while low temperature symmetry is possibly \textit{P2$_1$nm} as reported by several groups~\cite{Aladine, Wu, Jooss} using neutron diffraction and XRD data.

\begin{figure}
\includegraphics[width=8.5cm]{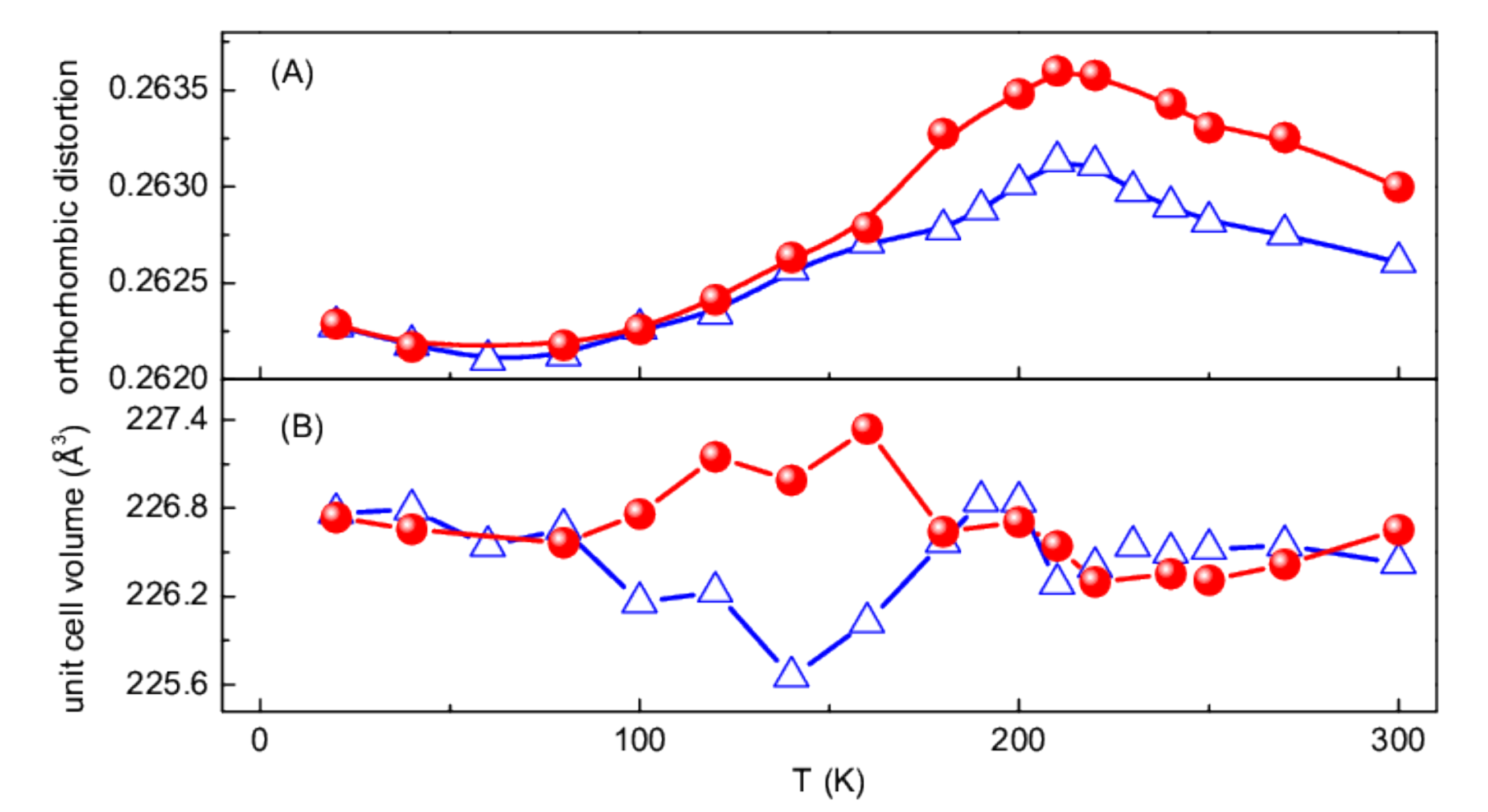}\\
\caption{(A) Plot of orthorhombic distortion versus temperature for the bulk PCMO1 and nanocrystalline PCMO3 showing that the orthorhombic distortion is reduced in PCMO3 compared to the bulk and that the distortion peaks near T$_{co}$. (B) Temperature dependence of the unit cell volume for both samples. The filled symbol correspond to the bulk PCMO1 while the open symbol represents PCMO3.}\label{fig:distortion}
\end{figure}

In case of nanoparticles it is difficult to conclude whether \textit{Pbnm} or \textit{P2$_1$nm} is the appropriate symmetry group irrespective of the temperature (Fig.~\ref{fig:spacegroup}). As predicted by several theoretical studies mentioned in the introduction, \textit{P2$_1$nm} space group symmetry (SGC) is non-centrosymmetric and hence there is a possibility of bulk PCMO being ferroelectric. We extracted lattice parameters a, b, c for PCMO1 and PCMO3 for the temperature range 20-300K and compared the evolution of orthorhombic distortion defined as $\Delta S=(a+b-c/\sqrt 2)/(a+b+c/\sqrt 2)$ with temperature. The orthorhombic distortion is unambiguously reduced in PCMO3 compared to the bulk. The temperature dependence of orthorhombic distortion, however, is qualitatively similar in both samples with a broad peak near $225$ K (Fig.~\ref{fig:distortion}A). Except for a small temperature range around $150$ K where marginal differences were observed, the unit cell volume remains unchanged for both systems (Fig.~\ref{fig:distortion}B). However, for PCMO4 the unit cell volume is considerably reduced (by more than 1\%) compared to other samples with higher crystallite size at room temperature (not shown in the figure). The reduction in orthorhombic distortion in nanocrystalline samples as compared to the bulk is expected because the crystal lattice is generally prone to develop towards a space group of higher symmetry as the particle size is reduced. There are several reports which gives strong evidence for structural phase coexistence between the ZP-CO/OO \textit{P2$_1$nm} and the charge and orbital disordered \textit{Pbnm} structure over a wide temperature range in bulk PCMO systems~\cite{Wu, Jooss}.  However, the qualitative similarity of the temperature dependence of lattice parameters in general and orthorhombic distortion in particular in PCMO1 and PCMO3 lead us to the conclusion that the non-centrosymmetric \textit{P2$_1$nm} phase may persist along with disordered \textit{Pbnm} phase at least down to crystal size of 55 nm.
\begin{figure}
\includegraphics[width=8.5cm]{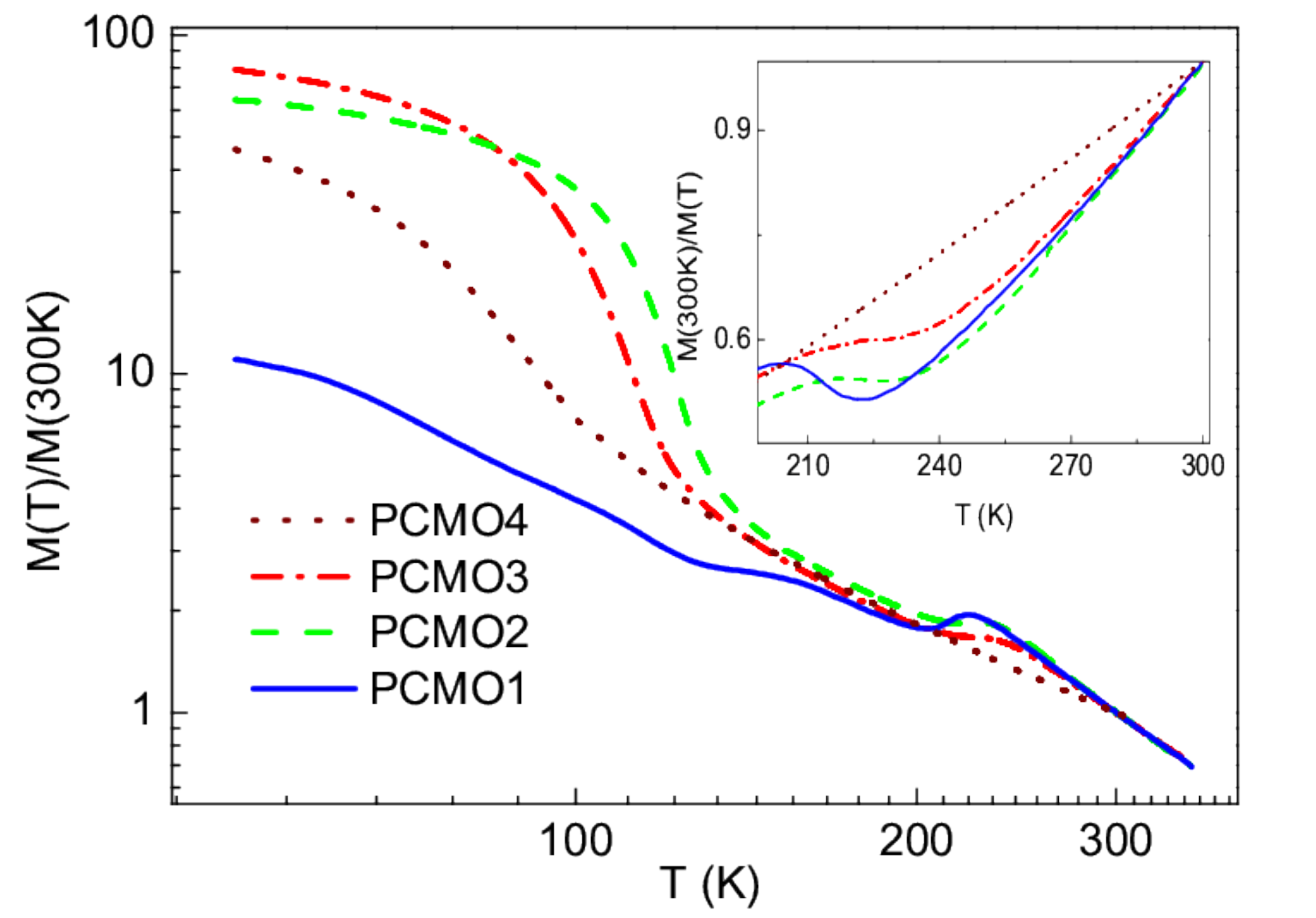}\\
\caption{Field cooled magnetization M(T) normalized with respect to M(300K) versus temperature (T) at magnetic field H=100 Oe for bulk PCMO1 (indicated by solid blue line) and nanocrystalline samples. The magnetic anomaly due to charge ordering is progressively broadened before vanishing entirely for PCMO4. Inset: Temperature dependence of the inverse of magnetization around $T_{co}$ for all samples, showing perfect Curie Weiss behaviour for PCMO4 which gives no evidence of ZP ordering.}\label{fig:mag}
\end{figure}

The temperature dependence of field cooled dc magnetization for four samples is presented in Fig.~\ref{fig:mag}. We observe an anomaly around T$_{co}$ in PCMO1, PCMO2 and PCMO3 while no such anomaly is observed in PCMO4. The magnetic anomaly is progressively broadened and eventually suppressed in PCMO4 as we decrease the particle size. The anomaly in the temperature dependence of magnetic susceptibility around $T_{co}$ can be explained within the framework of BC or ZP ordering with strong ferromagnetic correlations within the dimerised Mn pairs~\cite{Aladine} and weak ferromagnetic or AFM correlations between two such neighbouring pairs~\cite{Zheng}. The temperature dependence of inverse dc susceptibility for PCMO1, PCMO2 and PCMO3 deviates from the Curie Weiss law suggesting existence of ferromagnetic correlation even at room temperature (to be confirmed later). However, for PCMO4 we observe perfect Curie-Weiss behaviour with positive $\theta_{CW}\sim 74$K and with ZP type CO/OO completely suppressed. One can conclude that the \textit{P2$_1$nm} space group and hence its structural coexistence with the disordered \textit{Pbnm} phase is suppressed in PCMO4. The conclusion is in agreement with the previous works done by Ping Chai \textit{et al.}~\cite{Ping} and others~\cite{Fang, Samantaray, Giri}. On the low temperature side, the AFM ordering gives way to ferromagnetic like temperature dependence of susceptibility as the particle size is lowered. Dong \textit{et al.} attributed the ferromagnetic signal to the relaxation of super-exchange interaction on the surface layer leading to the formation of a ferromagnetic shell~\cite{Dong}. For PCMO4 the ferromagnetic signal is considerably reduced compared to other nanocrystalline samples possibly due to enhanced surface spin disorder and weakening of ferromagnetic correlations. It seems prima facie that if ZP type CO/OO is associated with spontaneous electric polarization~\cite{Colizzi}, PCMO4 should have reduced ZP induced ferroelectricity as compared to its bulk counterpart. On the other hand, the enhanced contribution of space inversion symmetry breaking near the grain boundaries cannot be ruled out as yet. The fact that the ZP ordering still persists and ferromagnetism enhanced in PCMO3 gives us the opportunity to study the particle size driven tunability of the coexistence of ferroelectricity and ferromagnetism.

\subsection{Impedance spectroscopy}

\begin{figure}
\includegraphics[width=8.5cm]{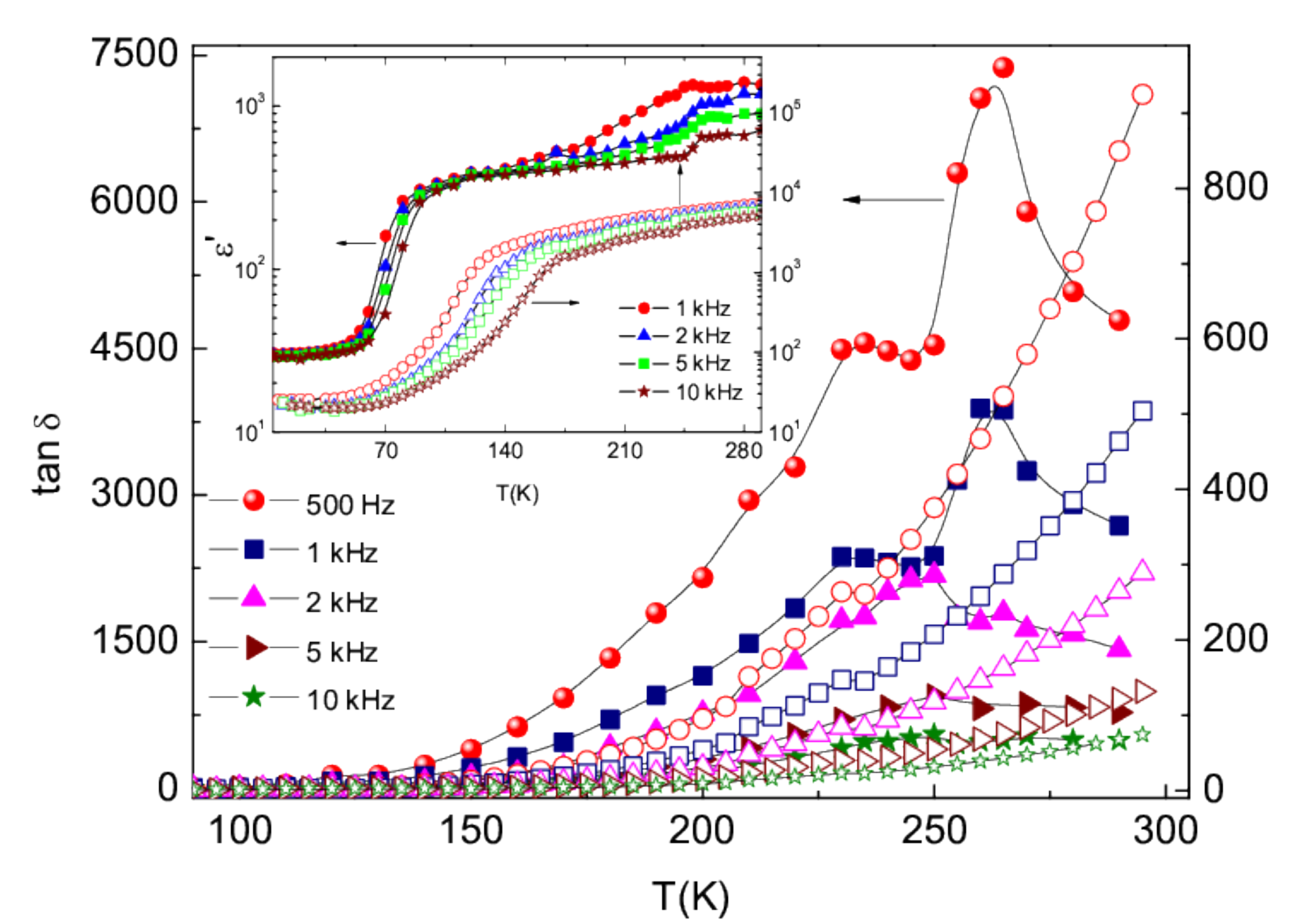}\\
\caption{The temperature dependence of the loss tangent for bulk PCMO1 as well as PCMO4: for PCMO1 the loss tangent shows two peaks at low frequency which merges into a single peak (at 240 K) at higher frequency whereas no peak in the loss tangent is observed for PCMO4 irrespective of the frequency regime. Inset: The temperature dependence of the real part of dielectric permittivity for PCMO1 shows an anomaly around 240 K which is absent in PCMO4. Filled and open symbols correspond to PCMO1 and PCMO4 respectively.}\label{fig:dielectric}
\end{figure}

The temperature dependence of real part of complex dielectric permittivity $\epsilon^{\prime}$ shows an anomaly around $240$ K for bulk sample while no such anomaly is observed for nanocrystalline systems (Inset, Fig.~\ref{fig:dielectric}). There is a sharp increase in the value of $\epsilon^{\prime}$ above $70$ K in qualitative agreement with reports by other groups~\cite{Freitas}. The dielectric anomalies have been attributed to Charge Density Wave (CDW) orderings or phase separation inhomogeneities before~\cite{Mercone, Freitas}. The dielectric loss tangent exhibits a distinct peak around the same temperature at high frequencies which splits into two as the frequency is lowered (Fig.~\ref{fig:dielectric}). Curiously, nanocrystalline samples do not show any peak in the loss tangent at any frequency within the measurable range and no anomaly in the temperature dependence of $\epsilon^{\prime}$ either. The large dielectric constant in PCMO cannot be attributed to the Schottky barriers at the sample-contact interface as nanocrystalline samples consistently show higher dielectric constant than their bulk counterpart suggesting the role of increased grain boundaries. The origin of the twin peaks in the temperature dependence of loss tangent in the lower frequency regime is unclear. After the merger of the twin peaks, the position of the loss tangent, however, do not shift with increase in frequency suggesting a phase transition. The absence of dielectric anomaly in case of nanocrystalline samples could be attributed to the broadening and the eventual suppression of ZP ordering transition.
\begin{figure}
\includegraphics[width=8.5cm]{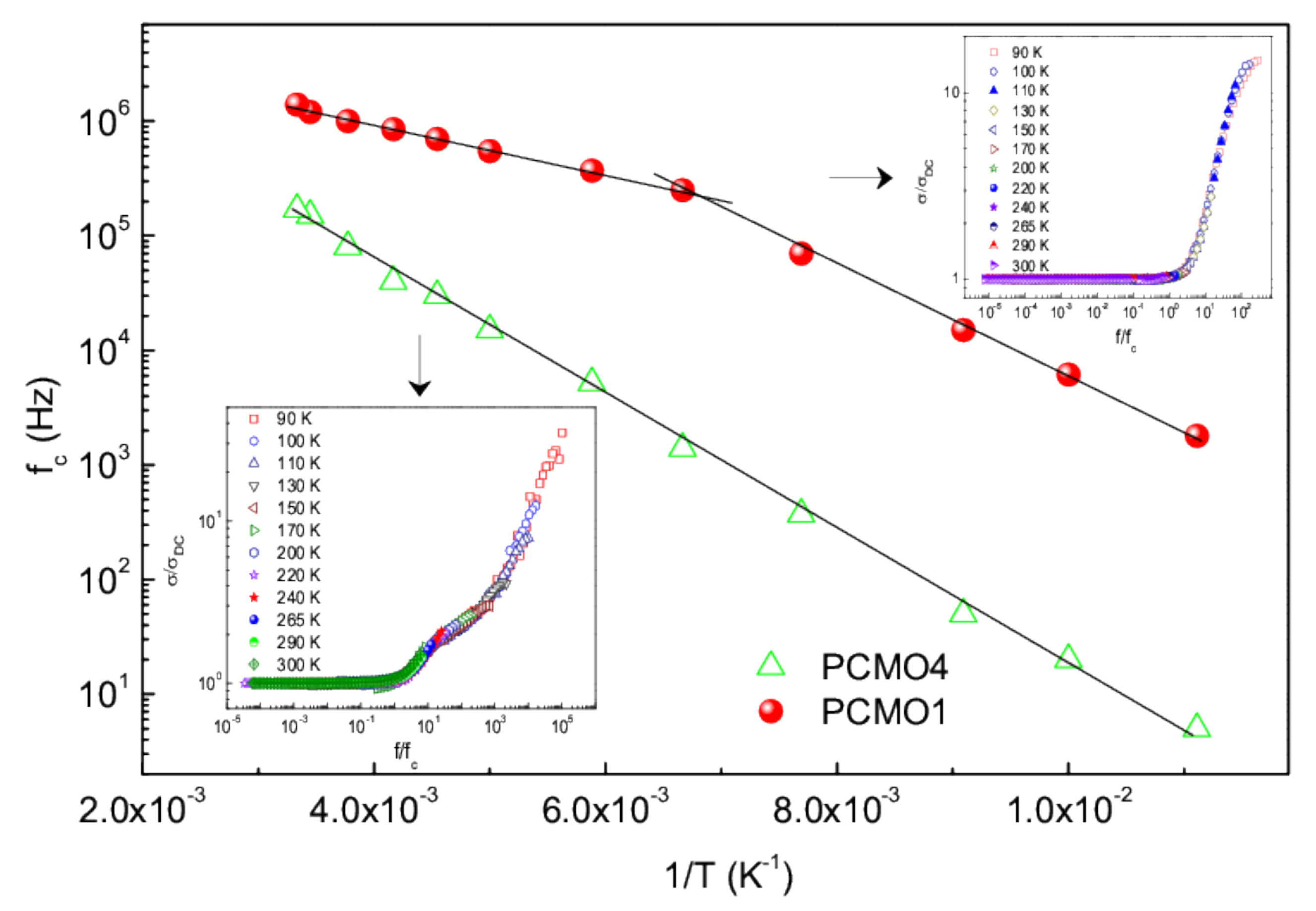}\\
\caption{The temperature dependence of the parameter $f_{c}$ obtained from the scaling of frequency dependence of ac conductivity at different temperatures for bulk PCMO1 (filled symbol) as well as nanocrystalline PCMO4 (open symbol). The universal scaling of the frequency dependence of conductivity $\sigma$ is shown for the bulk PCMO1 (Inset, top right) and nanocrystalline PCMO4 (Inset, bottom left).}\label{fig:hoppingrate}
\end{figure}

The frequency dependence of dielectric loss $\epsilon^{\prime\prime}$ shows power law scaling for PCMO1 and PCMO4 (not shown in the figure). Such behaviour, called the Universal Dielectric Response (UDR) is ubiquitous in dielectric systems and is generally associated with many body effect~\cite{Jonscher, Nagi}. The real part of complex conductivity $\sigma^{\prime}$ (henceforth to be denoted simply as $\sigma$) is related to the dielectric loss $\epsilon^{\prime\prime}$ as $\sigma(f)\sim f\epsilon^{\prime\prime}(f)$. The frequency dependence of $\sigma$ thus calculated at different temperatures for bulk as well as nanocrystalline samples collapses into a single master curve with an appropriate two parameter ($\sigma_{dc}$, $f_{c}$) scaling (Insets, Fig.~\ref{fig:hoppingrate}) defined by the following scaling relation
\begin{equation}
\sigma(f,T)= \sigma_{dc} G(\frac{f}{f_{c}}, T)
\end{equation}
This suggests that similar transport mechanism is prevalent throughout the temperature range of interest. The scaling parameter $f_{c}$ can be interpreted as the hopping rate of $e_{g}$ electrons or the critical frequency beyond which charge relaxation effect dominates. The temperature dependence of $f_{c}$ shows `activated' behaviour (the hopping rate increases exponentially with temperature) with bulk sample showing two distinct slopes indicative of two energy scales: $41.2\pm0.8$ meV below $140$ K (near the AFM ordering temperature) and $18.6\pm0.5$ meV above it (Fig.~\ref{fig:hoppingrate}). Interestingly the same plot for nanocrystalline samples (data for PCMO4 is shown in Fig.~\ref{fig:hoppingrate}) shows a single energy scale of $50.6\pm0.6$ meV which is close to the energy scale corresponding to lower $f_c$ regime in the bulk.
\begin{figure}
\includegraphics[width=8.5cm]{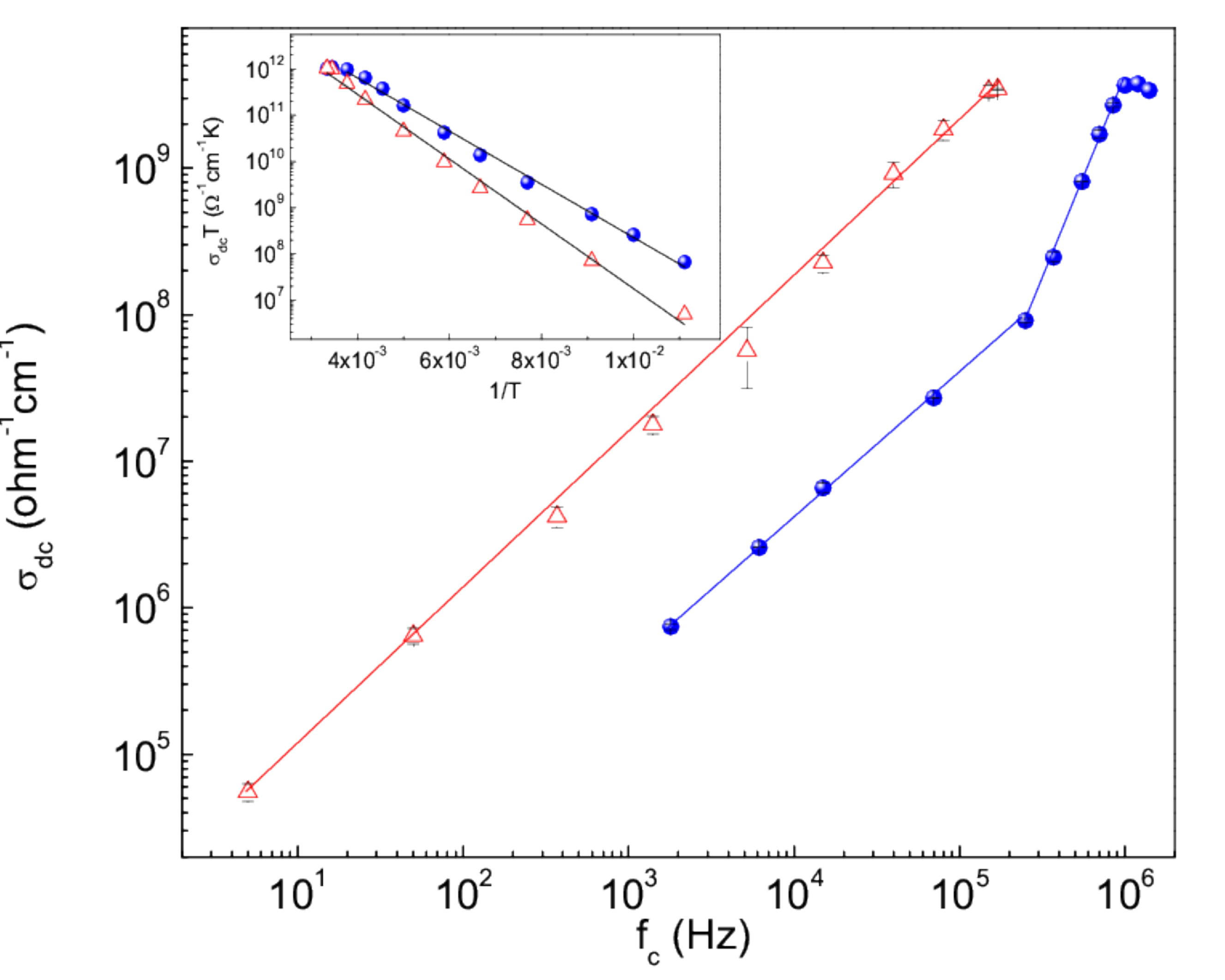}\\
\caption{The parameter $\sigma_{dc}$ extracted from the scaling of frequency dependent conductivity is plotted against critical frequency $f_{c}$ showing power law behaviour for both bulk PCMO1 (filled symbol) and nanocrystalline PCMO4 (open symbol). In the high frequency regime the power law exponent for the bulk is anomalously high. Inset: the temperature dependence of dc conductivity showing adiabatic small polaron hopping behaviour for PCMO1 and PCMO4.}\label{fig:powerlaw}
\end{figure}

The value of dc conductivity $\sigma_{dc}$ is calculated from the scaling of frequency dependent $\sigma$. A log-log plot of $\sigma_{dc}$ and $f_{c}$ shows power law behaviour with both $\sigma_{dc}$ and $f_{c}$ covering more than four decades of values. A least-square linear fit gives an exponent of $1.07\pm0.02$ for PCMO3 and $0.97\pm0.01$ for the bulk PCMO1 within the same $f_{c}$ regime (Fig.~\ref{fig:powerlaw}). An anomalously high value of $2.8$ in the high $f_{c}$ regime is observed for the bulk sample as well (Fig.~\ref{fig:powerlaw}). Interestingly the value of the exponent close to 1 is typical of binary disordered systems~\cite{Chakrabarty} and is quite justified taking into consideration the prevalence of electronic phase separation between conducting and insulating regions in hole doped mixed valence manganites. Nonetheless, the semi-log plot of $\sigma_{dc}T$ against $1/T$ (given in the inset) shows characteristics similar to adiabatic small polaronic transport with a single energy scale for bulk ($49.5\pm1.1$meV) as well as nanocrystalline ($60.4\pm1.5$meV for PCMO4) sample which are in agreement with the same calculated from dc resistivity measurements (not shown in the figure).

Let us now summarize the conclusions from impedance spectroscopy. In contrast to the bulk sample, nanocrystalline samples do not exhibit any signature of long range charge ordering usually characterized by the anomalies in the real part of dielectric permittivity and loss tangent. However the dielectric constant is significantly enhanced at room temperature compared to the bulk. The scaling of frequency dependent ac conductivity suggests similar transport mechanisms for bulk as well as nanocrystalline samples characteristic of small polaron hopping or that observed in binary disordered composite systems close to percolation threshold.

\subsection{Electron spin resonance spectroscopy}
There is a realistic possibility that a short-range charge-ordered state might exist when the long-range one is destroyed by the size reduction as observed before in the half-doped
manganite nanoparticles~\cite{Zhou, Zhou1}. Admittedly since charge ordering gets progressively weaker as one moves away from half-doping, one might expect complete suppression of even short range correlation below a certain critical size of the nanoparticle. Nevertheless, to establish the connection between ferroelectricity and ZP ordering it was necessary to investigate the existence or absence of short range spin charge correlations in PCMO4 where no long range ZP ordering has been observed according to the field cooled dc magnetic susceptibility, dc resistivity (not shown) and impedance spectroscopy data. In PCMO, the charge ordering is accompanied by orbital ordering in the ZP picture. Consequently even if the ordering is short ranged, any change in local micromagnetic configuration symmetry should be reflected either in the ESR spectra line width or the g-factor due to the change in spin-orbit coupling strength and the crystal field splitting.
\begin{figure}
\includegraphics[width=8.5cm]{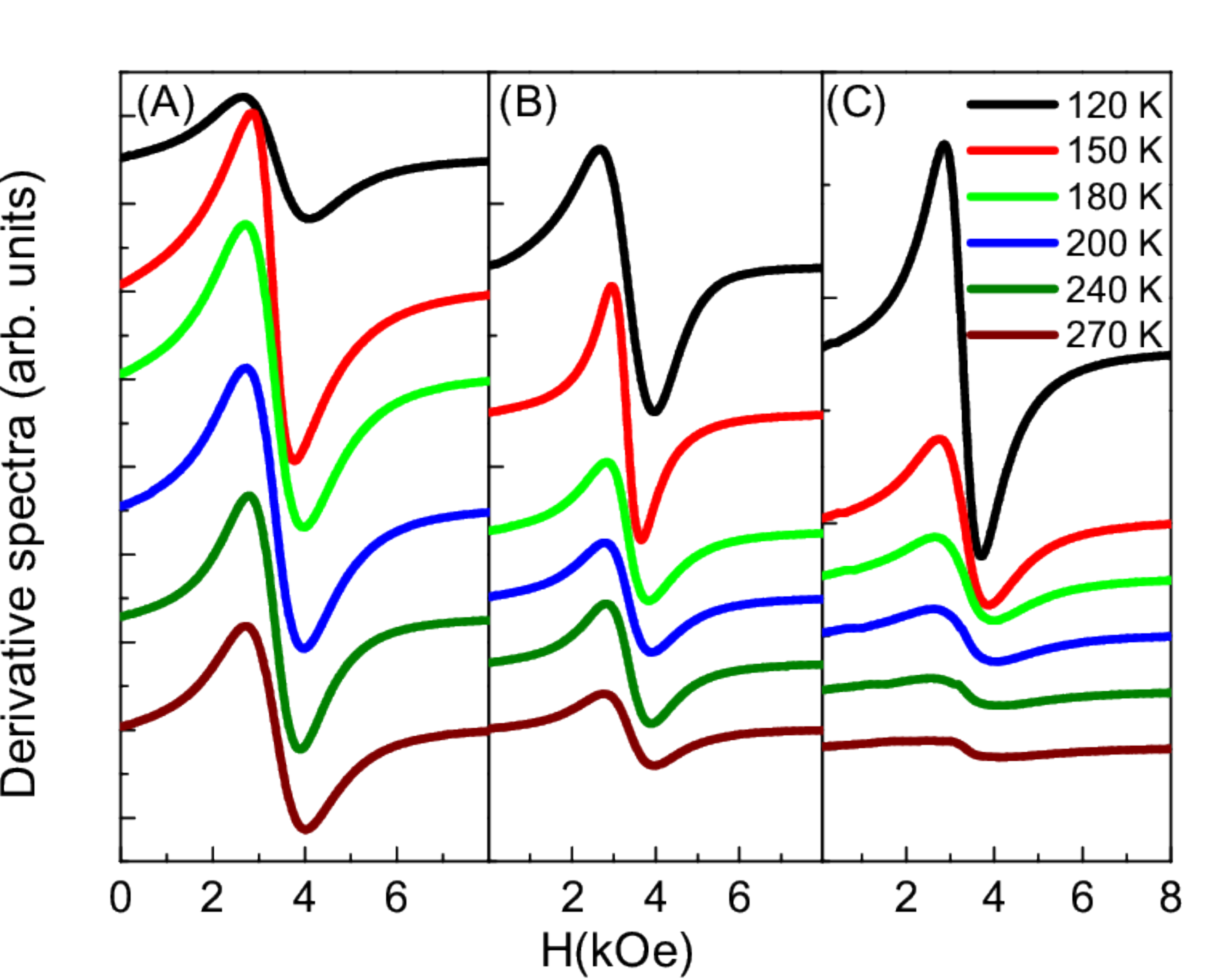}\\
\caption{A comparison of the ESR derivative spectra for PCMO1 (A), PCMO3 (B) and PCMO4 (C) at different temperatures. }\label{fig:esr0}
\end{figure}
\begin{figure}
\includegraphics[width=8.5cm]{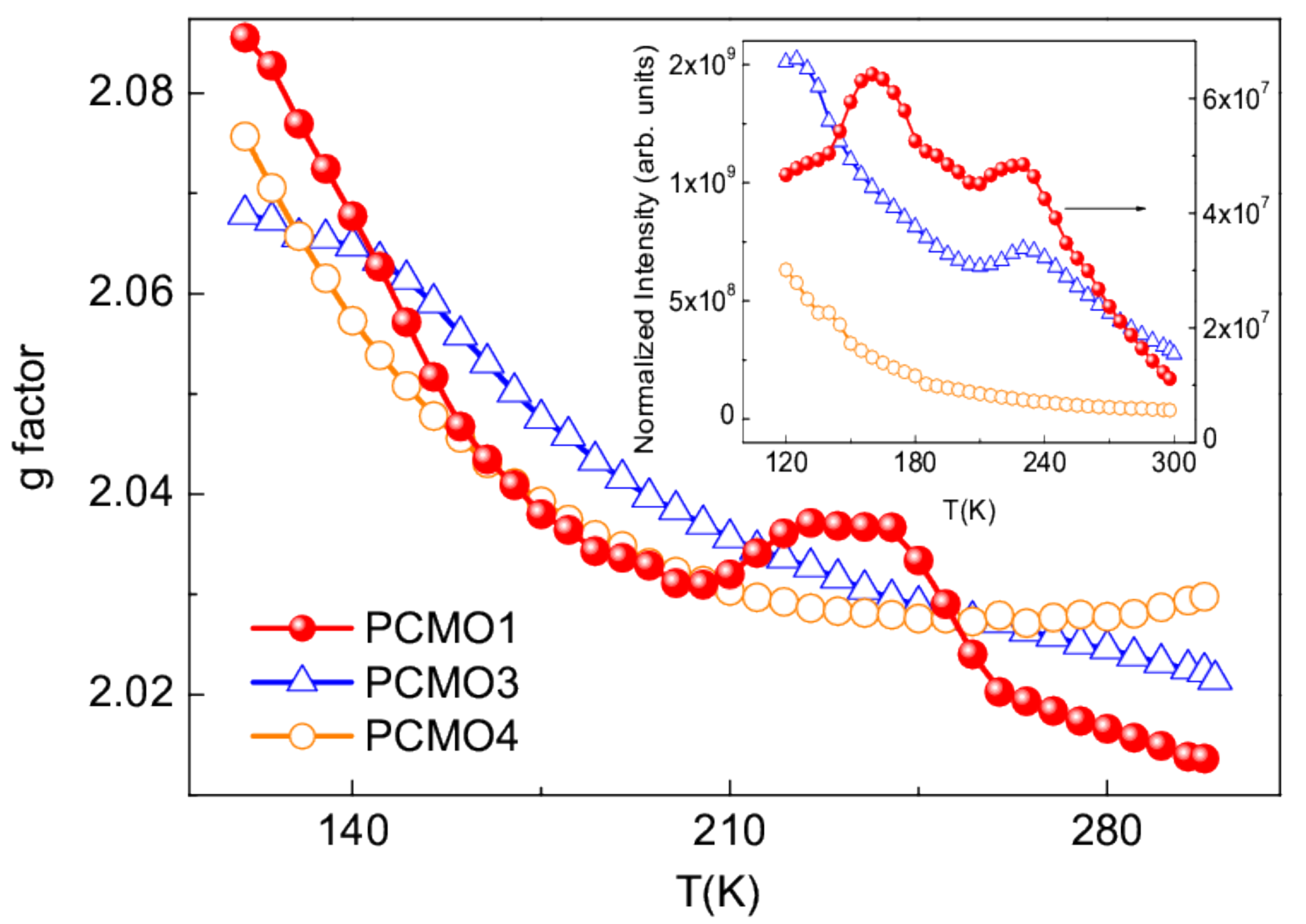}\\
\caption{The temperature dependence of the ESR g-factor for PCMO1, PCMO3 and PCMO4 is shown. A broad minimum is observed near room temperature for PCMO4 and an anomaly around $T_{co}$ for PCMO1. Inset: The temperature dependence of normalized intensity for all samples is consistent with field cooled dc susceptibility data discussed earlier.}\label{fig:esrgfactor}
\end{figure}
\begin{figure}
\includegraphics[width=8.5cm]{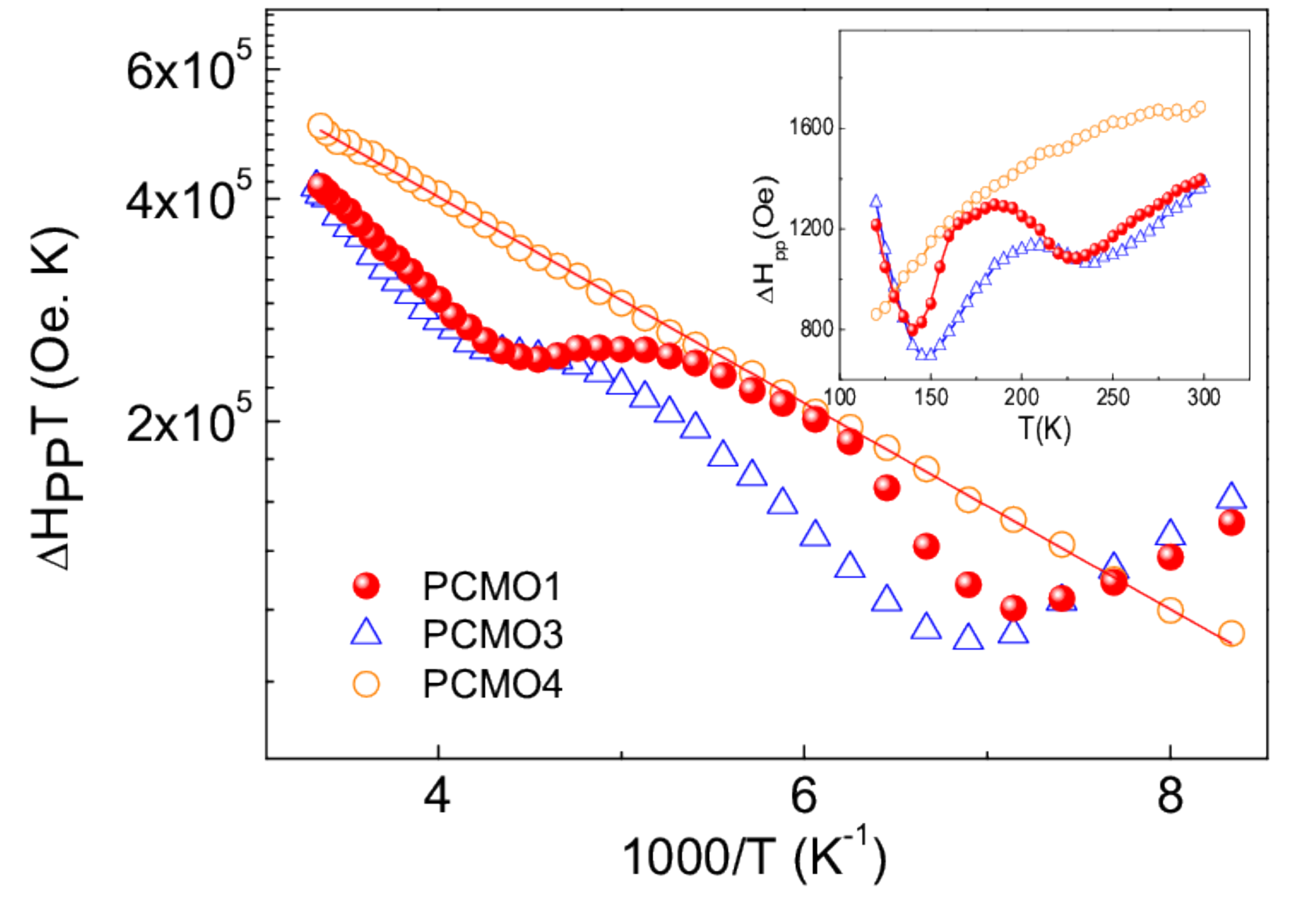}\\
\caption{The ESR line width (peak-to-peak) multiplied by temperature for PCMO4 is plotted against inverse of temperature which is similar to dc electrical conductivity or critical scaling frequency as discussed in the subsection for impedance spectroscopy. The continuous straight line is the linear fit to the data for PCMO4. Inset: The temperature dependence of ESR spectra line width showing the absence of short range correlation in PCMO4.}\label{fig:esrlinewidth}
\end{figure}

The ESR spectra of PCMO1, PCMO3 and PCMO4 at several representative temperatures within the range 110-300 K are displayed in the Fig.~\ref{fig:esr0}. No major difference between the samples was observed so far as line shape is concerned. For example, the spectrum at 270 K for all samples shows a broad single resonance line centered around $H_{r}\sim3.36$ kOe. However, we observe striking differences in the temperature dependence of $H_r$. For PCMO4, the resonance field first increases when cooled down from room temperature, shows a broad maximum followed by a decrease when temperature is lowered further (not shown in the figure) whereas for PCMO3 $H_r$ decreases monotonically with lowering of temperature.

The temperature dependent
intensity of the ESR absorption spectra, derived by integrating the first derivative spectra is shown in the Inset, Fig.~\ref{fig:esrgfactor} and is similar to the macroscopic magnetic susceptibility data. The temperature dependence of the g factor obtained from the resonance condition $h\nu=g\mu H_{r}$ is plotted for PCMO1, PCMO3 and PCMO4 in Fig.~\ref{fig:esrgfactor}. The minima in the g-factor for PCMO4 at 225 K marks the onset of short range ferromagnetic ordering, usually described within the framework of the `bottleneck model'~\cite{Shengelaya1} in which the Mn$^{4+}$ and Mn$^{3+}$ ions ferromagnetically
couple to form small polarons. This is indicated by the shift of resonance field to progressively lower value as the temperature is lowered. For PCMO1, the g-factor shows an anomaly around $T_{co}$. The anomaly is completely suppressed in PCMO3 which shows short range ferromagnetic ordering throughout the temperature range of interest. This is surprising since we observe clear anomalies around $T_{co}$ in the temperature dependence of dc susceptibility as well as ESR intensity for PCMO3 .

The temperature dependence of the spectral line width (peak-to-peak) calculated from the magnetic field derivative of the ESR absorption spectra exhibits a minimum close to $T_{co}$ and another minima around 140 K for PCMO1 and PCMO3 below which a sharp upturn in the line width is observed. For PCMO4, in sharp contrast to PCMO1 and PCMO3, the line-width decreases monotonically as the temperature is lowered from 300 K down to 110 K (Inset, Fig.~\ref{fig:esrlinewidth}). The broadening of the line-width above the minimum is attributed to the thermally activated hopping of small polarons. Below the minimum, the line width broadening
is associated with a critical `slowing down' of the spin fluctuations due to the diverging correlation length suggestive of a magnetic phase transition nearby.
\begin{figure}
\includegraphics[width=8.5cm]{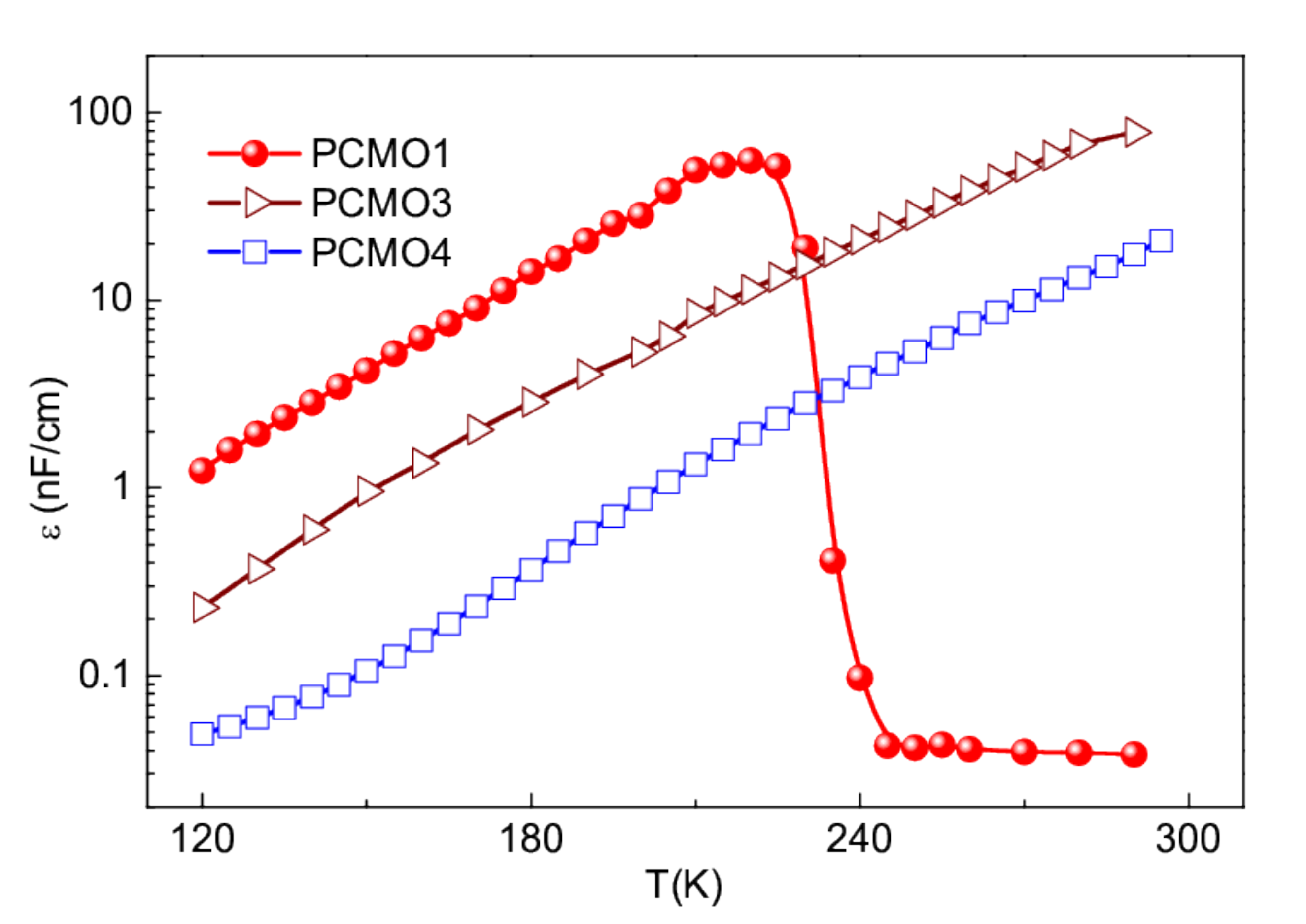}\\
\caption{The temperature dependence of the effective permittivity for bulk as well as nanocrystalline samples using PUND method. The applied electric field was 7 V/cm; a pulse width of 1ms and a delay time of 1000ms were used throughout the measurement. For bulk PCMO1, the permittivity shows sharp anomaly around the charge ordering temperature whereas no such feature is observed for nanocrystalline samples.}\label{fig:PUND1}
\end{figure}

Interestingly, Shengelaya \textit{et al.} showed that the ESR linewidth and electrical conductivity follow the same temperature dependence described by the small polaron hopping model in hole doped ferromagnetic manganites where spin lattice relaxation is mediated by the e$_{g}$ Jahn Teller polarons~\cite{Shengelaya2}. This is understandable as the conductivity in mixed valence manganites is determined by the probability of spin conserved $e_{g}$ electron hopping between nearest neighbouring sites while the broadening of the ESR linewidth arises due to the hopping rate of the $e_g$ electrons. As shown in Fig.~\ref{fig:esrlinewidth}, we find that in case of PCMO4, too, the ESR linewidth $\Delta H_{pp}$ and conductivity $\sigma$ show similar temperature dependence characteristic of the adiabatic hopping of small polarons albeit with much smaller value of the activation energy ($12.0\pm0.6$meV) in the former case. This result only reinforces our conclusion regarding short range FM correlations and no ZP ordering in PCMO4. The discrepancy in the energy scales obtained from temperature dependence of conductivity $\sigma$ or hopping rate $f_{c}$ and the ESR linewidth data for PCMO4 is strongly suggestive of separation of spin-charge relaxation. For PCMO3, which shows evidence of ZP ordering, the temperature dependence of $\sigma_{dc}$ (which does show evidence of adiabatic small polaron hopping throughout the temperature range of interest) and $\Delta H_{pp}$ is not identical. This suggests that the structural phase
coexistence between the ZP-CO/OO \textit{P2$_1$nm} and the disordered \textit{Pbnm} structure exists in PCMO3, too. The \textit{P2$_1$nm} phase being highly insulating, the electrical conduction is dominated by disordered \textit{Pbnm} phase. This allows the electrical transport in PCMO3 to be characteristic of adiabatic small polaron hopping whereas both \textit{P2$_1$nm} and \textit{Pbnm} contribute to the magnetic properties including ESR linewidth. In PCMO4, on the other hand, structural phase separation between \textit{P2$_1$nm} and disordered \textit{Pbnm} is completely suppressed.

Before we move on to the last segment of this article concerning the direct measurement of remanent electric polarization, let us summarize our conclusions from ESR spectroscopy. Field cooled magnetization together with ESR spectroscopy reveals that the magnetic correlation associated with charge ordering is completely suppressed as the particle size is reduced to $32$ nm. In the intermediate regime, we have an interesting situation where short range ferromagnetic correlations along with ZP ordering persists followed by long range ferromagnetic ordering at low temperature as opposed to the bulk which shows AFM ordering at low temperature. Moreover, the coexistence of \textit{P2$_1$nm} and disordered \textit{Pbnm} phase is completely suppressed in PCMO4.

\subsection{Measurement of dielectric permittivity and remanent electric polarization using PUND method}

The impedance spectroscopy was complemented by a study of temperature dependence of small signal capacitance and remanent polarization by measurement of the total charge on integrating the current through the capacitor using the Radiant Ferroelectric Tester. The problem of leakage associated with finite electrical conductivity in a polycrystalline pellet can be overcome by employing the Positive Up Negative Down (PUND) method~\cite{Naganuma}. Traditional P-E hysteresis loops are generally composed of ferroelectric, parasitic or stray capacitance and conductive contributions. PUND, also called the Double Wave method~\cite{Fukunaga}, measures remanent polarization by applying a set of three pulses: first, the preset pulse which sets the polarization of the sample in a certain direction and during which no measurement is made; second, a switching pulse is applied in the opposite direction of the preset pulse which reverses the sign of polarization and the total polarization including remanent and non-remanent part is measured; third, a non-switching pulse of same amplitude and duration is applied and the non-remanent polarization is measured; the difference between the two measurements gives us twice the remanent polarization. The time gap between the switching and non-switching pulse is called the delay time.
\begin{figure}
\includegraphics[width=8.5cm]{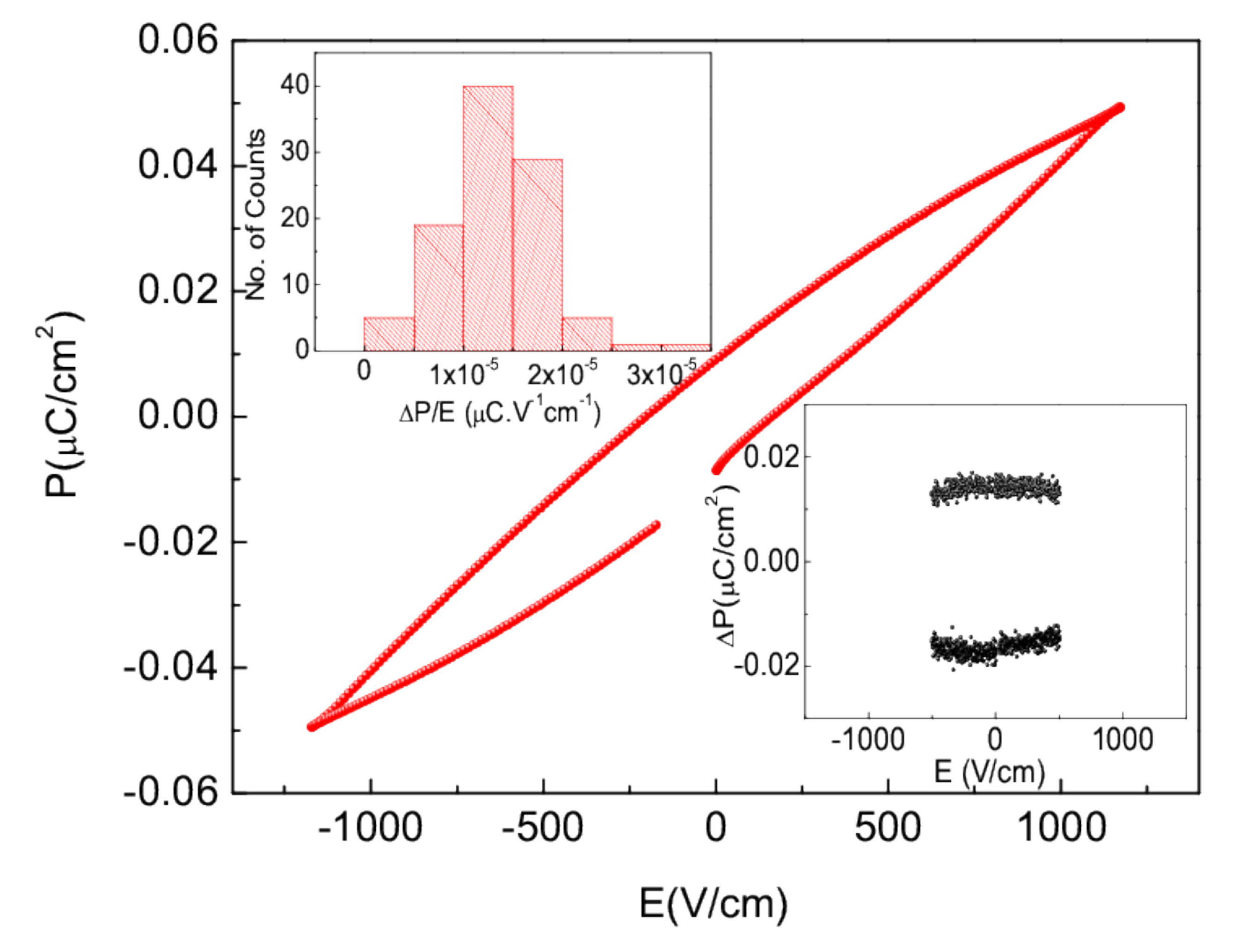}\\
\caption{The P-E hysteresis loop for PCMO3 at 77 K is shown. Inset (top left corner): The histogram of remanent polarization measured using PUND method at 77 K for PCMO3. The maximum applied electric field was 1.27 KV/cm and the pulse width of $40\mu$S. Inset (bottom right corner): The corresponding remanent polarization when subjected to the same maximum electric field using PUND method.} \label{fig:PUNDhyst}
\end{figure}

In case of specimens with relatively low resistivity in the temperature range of interest, the conduction current dominates over the relaxation current. The relaxation current can arise due to capacitative charging or slow switching of electric dipoles. On the other hand, for specimen with high resistivity, relaxation current dominates over the conduction current. That is why a choice of proper ratio between the pulse width and the delay time is important. Admittedly too large a delay time can also eliminate the effect of slow switching of electric polarization. Nevertheless, we chose a pulse width short enough to avoid dielectric breakdown and long enough to allow for slow switching of electric polarization to an extent and a long delay time $\sim$1000 ms to eliminate the effect of relaxation current. We also measured the dielectric permittivity from the effective capacitance calculated from the total charge (switching plus non-switching) per unit voltage pulse using PUND protocol and it turns out that the temperature dependence of dielectric permittivity shows a sharp drop around $T_{co}$ for bulk sample PCMO1 (Fig.~\ref{fig:PUND1}), far sharper compared to the anomaly in the real part of complex permittivity, the reason possibly being the difference in measurement technique, one calculated from measuring the impedance while for the other integrating the current output to a train of short-lived input pulses. For nanocrystalline samples PCMO3 and PCMO4, irrespective of whether the samples show any macroscopic signatures of ZP ordering or not, no such anomaly is observed (Fig.~\ref{fig:PUND1}), a result which is consistent with that obtained from impedance spectroscopy.
\begin{figure}
\includegraphics[width=8.5cm]{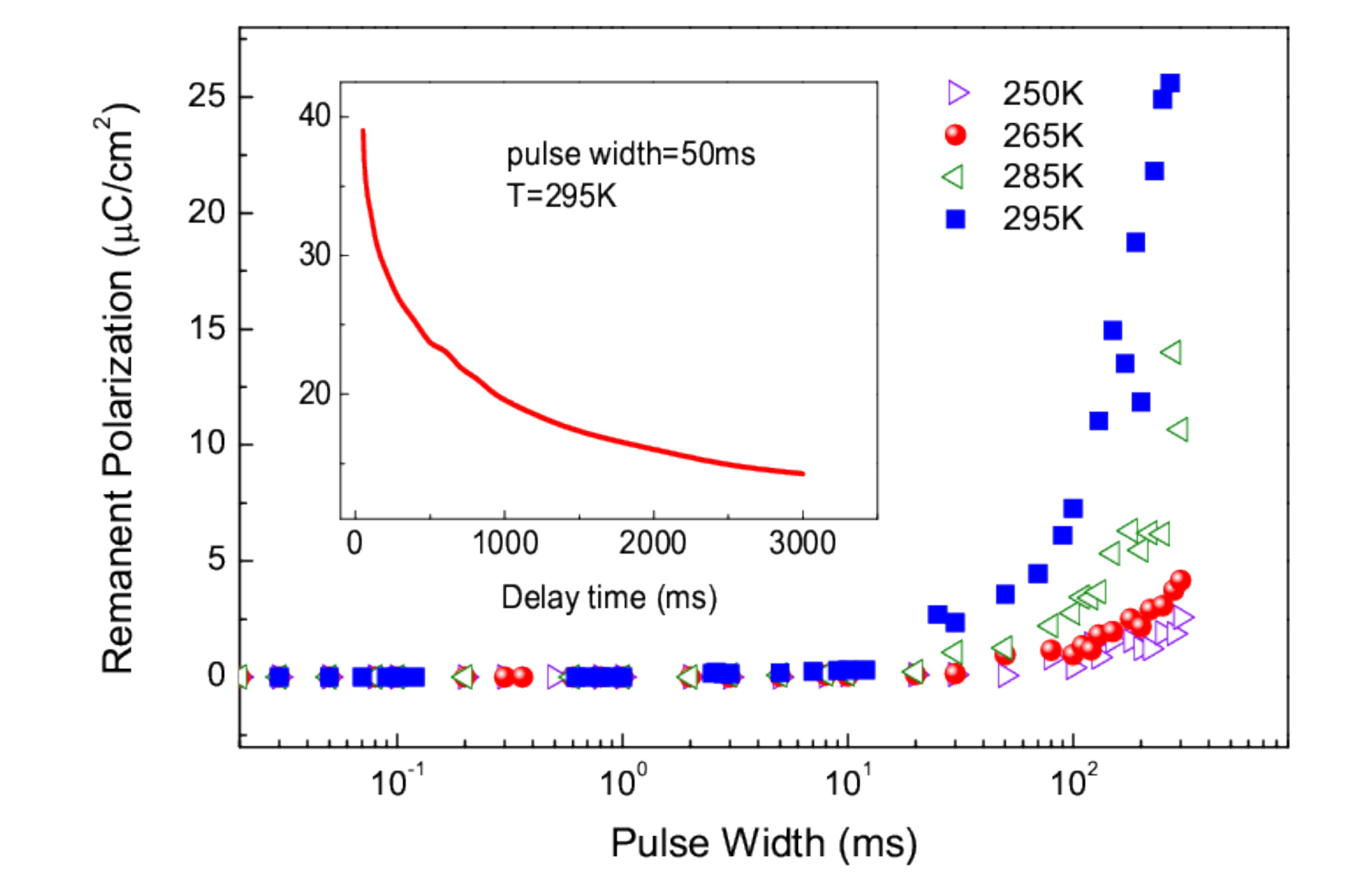}\\
\caption{The pulse width dependence of remanent polarization with applied electric field pulse 300 V/cm and delay time 3000 ms for different temperatures in PCMO3. The onset of switching occurs at progressively lower time scale as the temperature is increased. Inset: Delay time dependence of remanent electric polarization with applied field pulse 500 V/cm and pulse width 50 ms.}\label{fig:PUNDcontrol}
\end{figure}
\begin{figure}
\includegraphics[width=8.5cm]{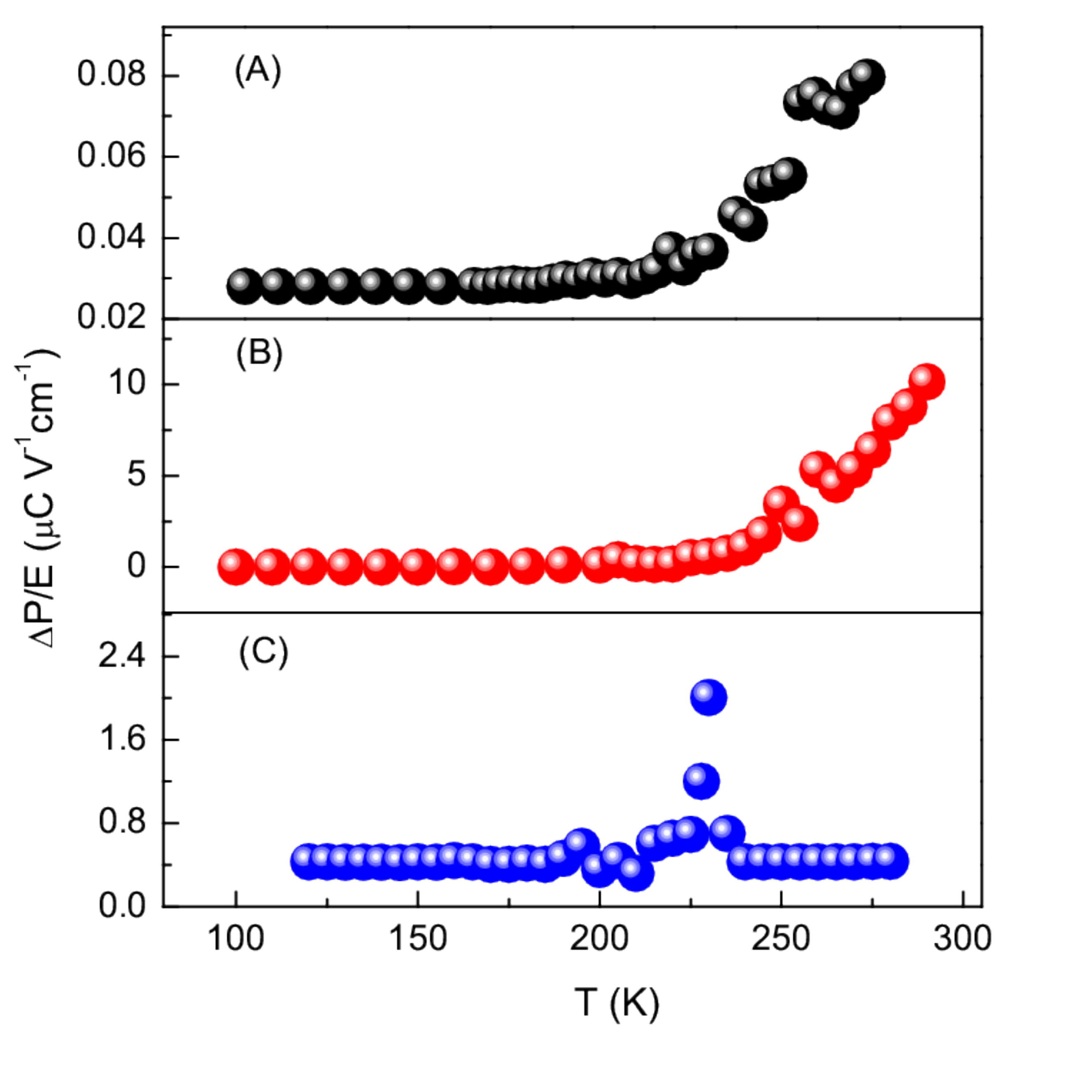}\\
\caption{The temperature dependence of the remanent polarization per unit applied field for bulk as well as nanocrystalline samples using PUND method. A pulse width of 1ms and a delay time of 1000ms were used throughout the measurement. C) An enhancement of ferroelectric polarization around the $T_{co}$ is visible in case of bulk PCMO1. B) For nanocrystalline samples, PCMO3 shows enhanced remanent polarization around room temperature;  A) The ferroelectric response is drastically suppressed in PCMO4 as the particle size is reduced to 32 nm.}\label{fig:remanent}
\end{figure}

The remanent polarization was measured using PUND method for pulses of varying time scale. It was observed that with increasing pulse width, above a critical value, the remanent polarization increases suddenly signalling the onset of polarization switching (Fig.~\ref{fig:PUNDcontrol}). With increasing temperature the critical time scale is reduced indicating that the switching is thermally activated. We did not find any saturation polarization as the pulse width cannot be increased indefinitely for a given delay time without letting other factors (discussed before) affect the electric polarization value.

Since the problem associated with leakage current is minimum at low temperature we measured the P-E hysteresis loop at 77 K without using PUND as shown in Fig.~\ref{fig:PUNDhyst}. The remanent polarization extracted from the corresponding hysteresis measurement obtained from PUND method (Inset, Fig.~\ref{fig:PUNDhyst}) matches well with the same calculated from the P-E loop. The switching is, however, stochastic in the sense that there is a distribution around the most probable value of remanent polarization (Inset, Fig.~\ref{fig:PUNDhyst}) if the same measurement is repeated a number of times (to be discussed in detail elsewhere). Despite the stochastic nature of polarization switching the temperature dependence of electric polarization (Fig.~\ref{fig:remanent}) does give important information. For the bulk sample we find an enhancement of electric polarization around T$_{co}$. The ferroelectric response is further magnified in case of PCMO3 where appreciable electric polarization is observed even at room temperature. We should keep in mind that PCMO3 shows signatures of ZP ordering and structural phase coexistence. Curiously, for PCMO4 which does not show any evidence of ZP ordering / structural coexistence, the electric polarization is even lower than the bulk. This is a strong indicator for the connection between ZP ordering / structural phase coexistence and ferroelectricity. The result cannot be attributed to leakage problem related to finite electrical conductivity as the resistivity near room temperature becomes progressively higher as the particle size is lowered and the electric polarization value in different samples has little correlation with the corresponding value of resistance.

\section{Summary}
We have investigated the possible connection between ZP ordering and ferroelectricity in hole doped manganites using temperature dependent XRD scans, magnetization measurement, impedance spectroscopy, direct measurement of remanent electric polarization using PUND method and ESR spectroscopy. The Rietveld structural analysis of the temperature dependent XRD scans prima facie indicate that the high temperature phase of bulk sample is centrosymmetric \textit{Pbnm} with the non-centrosymmetric \textit{P2$_1$nm} phase associated with Zener polaron ordering dominating at low temperature. The orthorhombic distortion is progressively reduced with lowering of crystallite size but no change in space group symmetry was observed in nanocrystalline samples as compared to the bulk. The magnetic anomaly associated with ZP ordering gets progressively suppressed as the particle size is lowered while the AFM ordering in the bulk at low temperature is replaced by ferromagnetic like ordering in nanocrystalline systems. A closer inspection using ESR spectroscopy finds evidence of short range ferromagnetic correlations in nanocrystalline samples near room temperature as well.

The impedance spectroscopy shows an anomaly in the temperature dependence of real part of complex dielectric permittivity for bulk PCMO1. This is accompanied by a peak in the loss tangent around T$_{co}$ suggesting ferroelectric phase transition. Such behaviour is completely suppressed in nanocrystalline samples. On the other hand the dielectric permittivity is enhanced by orders of magnitude at room temperature on lowering the crystallite size. The scaling of the frequency dependent conductivity and power law dependence of dc conductivity on the scaling frequency suggests similar transport mechanism in bulk as well as nanocrystalline samples, typical of binary disordered composite systems. The temperature dependence of dc conductivity in nanocrystalline as well as bulk sample shows characteristics similar to adiabatic small polaron hopping through the backbone of disordered \textit{Pbnm} phase. The dissimilarity between the temperature dependence of $\sigma_{dc}$ and $\Delta H_{pp}T$ persists down to very low particle size, which suggests prevalence of structural phase separation between \textit{P2$_1$nm} and disordered \textit{Pbnm} over a wide temperature range. However, below certain critical size, the temperature dependence of $\sigma$ and $\Delta H_{pp}$ is strikingly similar indicating complete suppression of ZP ordering and accompanying structural phase coexistence.

The direct measurement of dielectric permittivity and remanent electric polarization using PUND method not only reproduces the major results from impedance measurements but establishes that the remanent electric polarization can be significantly enhanced near room temperature in nanocrystalline samples provided ZP ordering is not suppressed. The stochasticity or non-deterministic behaviour of the ferroelectric response is, however, not particularly attractive from the applications point of view. Although the observed ferroelectric response is not robust, as a welcome byproduct of the exercise we have at least the tangible possibility of short range ferromagnetic correlation and ferroelectricity coexisting at room temperature paving way for further scope of research in this direction.

In a nutshell, we present direct experimental evidence that ferroelectric response in manganites is indeed related to ZP ordering and accompanying structural phase separation as predicted theoretically. There is also a distinct possibility of not just enhancing ferroelectric response but combining ferroelectric and ferromagnetic correlations in nanocrystalline ZP ordered, phase separated manganites near room temperature.

\section{Acknowledgements}
VKS and KD thanks CSIR, India for providing financial support.

\end{document}